\documentclass{svjour3}                     
\smartqed  
\usepackage{graphicx}
\usepackage{amsmath,amsfonts,amsbsy,amssymb,fullpage,makeidx,array,graphicx,epsfig,enumerate}

 \newcommand{\df}{\ {\overset {\rm def} =}\ }
 \newcommand{\dr}[2]{\frac {{\rm d} {#1}} {{\rm d} {#2}}}
 
 \newcommand{\dril}[2]{{{\rm d} {#1}} / {{\rm d} {#2}}}
 \newcommand{\pdril}[2]{{\partial {#1}} / {\partial {#2}}}
 
 \newcommand{\llim}[1] {\ {\underset {#1} {\longrightarrow}}\ }

\newcommand{\pb}[2]{
\parbox[t]{#1}{\raggedright #2}
}

\newcommand{\td}[2]{\frac{{\rm d} {#1}}{{\rm d} {#2}}}

\newcommand{\pdil}[2]{\partial {#1} / \partial {#2}}

\renewcommand{\th}{\widehat{t}}
\newcommand{\Rh}{\widehat{R}}
\newcommand{\er}[1]{(\ref{#1})}
\begin{document}

\title{Imitating accelerated expansion of the Universe by matter
inhomogeneities
-- \\
corrections of some misunderstandings}

\titlerunning{Misunderstandings about accelerated expansion}

\author{Andrzej Krasi\'nski
  \and
  Charles Hellaby
  \and
  Krzysztof Bolejko
  \and
  Marie-No\"elle C\'el\'erier
}

\institute{Andrzej Krasi\'nski \\
N. Copernicus Astronomical Centre, Polish Academy of Sciences,
Bartycka 18, 00 716 Warszawa, Poland, \\
           email: akr@camk.edu.pl \\
 \and
Charles Hellaby, \\
Department of Mathematics and Applied Mathematics,
University of Cape Town, Rondebosch 7701, South Africa, \\
           email: Charles.Hellaby@uct.ac.za \\
           \and
Krzysztof Bolejko, \\
N. Copernicus Astronomical Centre, Polish Academy of
Sciences, Bartycka 18, 00 716 Warszawa, Poland \\
      \and
Department of Mathematics and Applied Mathematics,
University of Cape Town, Rondebosch 7701, South Africa, \\
           email: bolejko@camk.edu.pl \\
 \and
Marie-No\"elle C\'el\'erier, \\
Laboratoire Univers et Th\'eories (LUTH), Observatoire de Paris, CNRS,
Universit\'e Paris-Diderot,
5 place Jules Janssen, 92190 Meudon, France, \\
           email: marie-noelle.celerier@obspm.fr
}

\maketitle

\begin{abstract}
A number of misunderstandings about modeling the apparent accelerated expansion
of the Universe, and about the `weak singularity' are clarified: 1. Of the five
definitions of the deceleration parameter given by Hirata and Seljak (HS), only
$q_1$ is a correct invariant measure of acceleration/deceleration of expansion.
The $q_3$ and $q_4$ are unrelated to acceleration in an inhomogeneous model. 2.
The averaging over directions involved in the definition of $q_4$ does not
correspond to what is done in observational astronomy. 3. HS's equation (38)
connecting $q_4$ to the flow invariants gives self-contradictory results when
applied at the centre of symmetry of the Lema\^{\i}tre--Tolman (L--T) model. The
intermediate equation (31) that determines $q_{3'}$ is correct, but approximate,
so it cannot be used for determining the sign of the deceleration parameter.
Even so, at the centre of symmetry of the L--T model, it puts no limitation on
the sign of $q_{3'}(0)$. 4. The `weak singularity' of Vanderveld {\it et al.} is
a conical profile of mass density at the centre -- a perfectly acceptable
configuration. 5. The so-called `critical point' in the equations of the
`inverse problem' for a central observer in an L--T model is a manifestation of
the apparent horizon -- a common property of the past light cones in zero-lambda
L--T models, perfectly manageable if the equations are correctly integrated.
\end{abstract}

\maketitle

\section{Motivation}

Vanderveld {\it et al.} \cite{VFWa2006} (abbreviated as VFW) claimed that there
was a contradiction between two results concerning the putative accelerated
expansion of the Universe. Their reasoning was, in brief, this (italics mark
quotations): On the one hand, Hirata and Seljak \cite{HiSe2005} (abbreviated as
HS) claimed to have proved that in a perfect fluid cosmological model that is
geodesic, rotation-free and obeys the strong energy condition $\rho + 3p \geq
0$, a certain generalisation of the deceleration parameter $q$ must be
non-negative. But on the other hand, Iguchi {\it et al.} \cite{INNa2002}
(hereafter INN) did obtain simulated acceleration in Lema\^{\i}tre -- Tolman
(L--T) models \cite{Lema1933,Tolm1934} with $\Lambda = 0$ that obey HS's
conditions. This {\em contradiction} is resolved by showing that L--T models
that simulate accelerated expansion also contain a {\em weak singularity}, and
in this case the derivation of HS breaks down. In addition to this, there are
\textsl{other singularities that tend to arise} in L--T models, and VFW
\textsl{have failed to find any singularity-free models that agree with
observations.}

This leaves the impression that physically acceptable inhomogeneous models are
unable to account for observations. It is shown here that this reasoning is not
correct. In brief, our theses are the following:

1. Of the five definitions of the deceleration parameter given by HS, only $q_1$
is a correct invariant measure of \noindent deceleration of fluid expansion.
Their $q_{3'}$ derives from a Taylor expansion of the luminosity
distance--redshift relation, $D_L(z)$, in powers of $z$, done separately for
each direction. Their $q_4$ is based on the angular average of a Taylor
expansion of $z(D_L)$, and is conceptually different from what is done in
observational practice. Although all these definitions reduce to the familiar
$q$ in FLRW models, they have different values and distinct meanings in an
inhomogeneous model. In particular $q_{3'}$ and $q_4$ are not measures of
deceleration in an inhomogeneous model.

2. HS's $q_4$ does not correctly represent the observers' deceleration parameter
$q$. When observers assume that our real Universe is in the FLRW class, $q$ is
the same in all directions. But when they consider inhomogeneous and anisotropic
models, they have to measure the distance--redshift relation for each direction
separately rather than average the measured results over directions, as this
would mean destroying useful information. Thus, HS's $q_4$ that represents the
result of such averaging does not correspond to the observations actually done
in relation to this quantity.

3. HS's reasoning is correct (in the approximate sense) up to their eq. (31),
which allows one to calculate $q_{3'}$ at the centre of symmetry in a
Lema\^{\i}tre -- Tolman (L--T) model. But, at a spherical centre, the final
result is the opposite to what they intended -- no limitation on the sign of
$q_4(0)$ follows (see our Secs. \ref{HSproblems} and \ref{Fproblems}).

Thus, there is no contradiction involved in a decelerating inhomogeneous model
imitating observational relations of an accelerating FLRW model. Even so, it is
incorrect to use an approximate equation to decide whether some quantity is
positive or negative.

4. What VFW call a {\em weak singularity} is not a singulalrity.
\footnote{Moreover, their definition of `weak singularity' does not agree with
the meaning of the term `singularity' in relativity, and is not supported by
anything in the Tipler paper \cite{Tipl1977} they cite.}

5. It is also not true that \textsl{other singularities} invalidate the L--T
models. The differential equations for the \textsl{inverse problem} -- given
observational data functions, calculate the LT model that would give them -- are
indeed singular at the apparent horizon. VFW refer to this location as a
\textsl{critical point}, and to this phenomenon as a \textsl{pathology}. In
truth, this is simply the reconvergence of the observer's past null cone towards
the Big Bang, that is a consequence of the decelerating cosmic expansion when
lambda is zero, long known in the FLRW case, e.g \cite{Elli1971,Hoyl61,McC34}.
Though some authors (INN and VFW among them) were unable to propagate their
solutions through the `critical point', this difficulty was overcome by Lu and
Hellaby \cite{LuHe2007} and in fact the AH relation was used to provide extra
information by McClure and Hellaby \cite{Hell2006,McHe2008}.

Since these misunderstandings have propagated into the literature, it is
essential to resolve these issues.

But the most important point to be stressed is this: what one wants to
reproduce, in connection with the distant type Ia supernovae, is not the
accelerated expansion of the Universe, but the observed luminosity
distance--redshift relation. The apparent dimming of supernovae (first reported
by Riess {\it et al.} \cite{Riess1998} and Perlmutter {\it et al.}
\cite{Perl1999}) stems from a comparison between observations and the
Einstein-de Sitter model, the `old' standard model of the Universe. The first
proposed explanation, originating from the assumption that the Universe should
be Friedmannian, is an accelerated expansion. However, this acceleration is not
the phenomenon to be explained, but a component of a particular explanation --
one that assumes homogeneity on the scale in question.

\section{The Lema\^{\i}tre -- Tolman (L -- T) model}

Since in the following we often refer to the L--T model, we summarise here the
basic facts about it in the case $\Lambda = 0$. For more extended expositions
see Refs. \cite{Kras1997,PlKr2006}. The metric of the L--T model is:
\begin{equation}\label{2.1}
{\rm d} s^2 = {\rm d} t^2 - \frac {{R_{,r}}^2}{1 + 2E(r)}{\rm d} r^2 -
R^2(t,r)({\rm d}\vartheta^2 + \sin^2\vartheta \, {\rm d}\varphi^2),
\end{equation}
where $E(r)$ is an arbitrary function, and $R(t, r)$ is determined by the
integral of the Einstein equations:
\begin{equation}\label{2.2}
{R_{,t}}^2 = 2E(r) + 2M(r) / R,
\end{equation}
$M(r)$ being another arbitrary function. Equation (\ref{2.2}) has the same
algebraic form as one of the well-known Friedmann equations, except that here it
contains arbitrary functions of $r$ in place of arbitrary constants. The
Friedmann limit follows when $E = - k r^2 / 2, M = M_0 r^3$ and $R = r S(t)$
where $k$, $M_0$ and $S(t)$ are the corresponding Friedmann constants and scale
factor. The solution of (\ref{2.2}) may be written as
\begin{equation}\label{2.3}
t - t_B(r) = \int \frac {{\rm d} R} {\pm \sqrt {2E(r) + 2M(r) / R}},
\end{equation}
where $t_B(r)$ is one more arbitrary function called the bang-time function; in
the Friedmann limit it is constant. The $+$ sign applies for an expanding
region, $-$ applies for a collapsing region. The mass density is
\newpage
\begin{equation}  \label{2.4}
8\pi G \rho = \frac {2{M_{,r}}}{R^2R_{,r}}.
\end{equation}
The pressure is zero, and so the matter (dust) particles move on geodesics.

The equations determining the L--T model are covariant with the transformations
$r \to r' = f(r)$. Therefore, we may use such a transformation to give one of
the functions $(M, E, t_B)$ a handpicked form, provided the chosen function is
monotonic in the range under investigation.

An incoming radial null geodesic is given by the differential equation
\begin{equation}\label{2.5}
\dr t r = - \frac {R_{,r}} {\sqrt{1 + 2E(r)}},
\end{equation}
and its solution is denoted $t = \widehat{t}(r)$. The general solution
determines the past null cone (PNC) of the observer situated at the centre of
symmetry. We use the hat $\,\widehat{}\,$ and the subscript $\wedge$ to indicate
evaluation on the PNC. Then the redshift along the PNC, $z(r)$, is given by
\cite{Bond1947,PlKr2006}:
\begin{equation}\label{2.6}
\frac 1 {1 + z}\ \dr z r = \left[ \frac {R_{,tr}} {\sqrt{1 + 2E}} \right]_\wedge
~,
\end{equation}
and the area and luminosity distances are
\begin{align}\label{2.7}
 D_A = \Rh,~~~~~~ D_L = (1 + z)^2 \Rh.
\end{align}
Equations (\ref{2.5}) and (\ref{2.6}) can be solved numerically once $M(r)$,
$E(r)$ and $t_B(r)$ are specified.

An L--T model may possibly have a curvature singularity at the centre of
symmetry. The singularity will be absent when there is no point mass there, and
the mass density is finite. This will happen when the conditions given below are
obeyed (see Ref. \cite{PlKr2006} for the derivation).

Let $r = r_c$ be the radial coordinate of the centre of symmetry, where $R(t,
r_c) = 0$ for all $t > t_B(r_c)$, and let $\rho(t, r_c) \df \alpha(t) < \infty$.
We choose the $r$-coordinate so that $r_c = 0$ and
\begin{equation}\label{2.8}
M = M_0 r^3,
\end{equation}
where $M_0 =$ constant. Then, in a neighbourhood of $r = 0$, the functions
behave as follows:
\begin{equation}\label{2.9}
R = \beta(t) r + {\cal R}_1(r), \quad E = \gamma r^2 + {\cal R}_2(r), \quad t_B
= \tau + {\cal R}_0(r),
\end{equation}
$\gamma$ and $\tau$ being constants, where the symbols ${\cal R}_a(r)$ will
denote quantities with the property
\begin{equation}\label{2.10}
\lim_{r \to 0} \frac {{\cal R}_a(r)} {r^a} = 0.
\end{equation}

\section{The definitions of deceleration by HS and their physical meaning}
\label{4def}

\setcounter{equation}{0}

Hirata and Seljak \cite{HiSe2005} give in their paper five alternative
definitions of the deceleration parameter, three of which we briefly recall here
for reference. In all cases it is assumed that the spacetime is filled with a
perfect fluid obeying Einstein's equations with zero cosmological constant. HS
assume that the fluid moves geodesically, but for the beginning we shall drop
this assumption.

Their first definition of the deceleration parameter is:
\begin{equation}\label{3.1}
q_1 \df - 1 - u^{\mu} H_{1; \mu} / {H_1}^2, \qquad H_1 \df {u^{\mu}}_{; \mu}/3,
\end{equation}
where $H_1$ is the (local) Hubble parameter and $u^{\mu}$ is the velocity field
of the fluid. Then, using these definitions, the Raychaudhuri equation may be
rewritten as follows:
\begin{equation}\label{3.2}
{H_1}^2 q_1 = \frac 1 3 \left[- {\dot{u}^{\gamma}}{}_{;\gamma} + \sigma^{\mu
\nu} \sigma_{\mu \nu} - \omega^{\mu \nu} \omega_{\mu \nu}\right] + \frac {4 \pi
G} 3 (\rho + 3p),
\end{equation}
where $\dot{u}^{\mu}$ is the acceleration of the fluid, $\sigma_{\mu \nu}$ and
$\omega_{\mu \nu}$ are the shear and rotation tensors, respectively, $\rho$ is
the mass density and $p$ is the pressure. It follows that with $\dot{u}^{\mu} =
0 = \omega_{\mu \nu}$ and $\rho + 3p \geq 0$, $q_1$ must be non-negative. This
result applies in particular to the L--T models (where $\dot{u}^{\mu} = 0 =
\omega_{\mu \nu}$ and $p = 0$) and to the FLRW models (where $\dot{u}^{\mu} = 0
= \omega_{\mu \nu} = \sigma_{\mu \nu}$).

The definition (\ref{3.1}) is invariant, applies in all geometries, is a
generalization of the deceleration parameter long used in the FLRW geometries,
and measures the relative change of the expansion scalar along the flow lines of
the perfect fluid. With $q_1 \geq 0$ we have $u^{\mu} H_{1; \mu} \leq - {H_1}^2
< 0$, so the expansion slows down toward the future of the comoving observers.
This is the case in all cosmological models that obey the assumptions listed
above.

The quantities $q_{3'}$ and $H_{3'}$ refer to a Taylor expansion of the
luminosity distance-redshift relation $z(D_L)$:
\begin{align}\label{3.3}
z & = H_{3'} D_L - \frac 1 2 {H_{3'}}^2 (1 - q_{3'}) {D_L}^2 + {\cal
O}({D_L}^3),
\end{align}
where $H_{3'}$ and $q_{3'}$ are functions of the angle of observation. The
definitions of $H_4$ and $q_4$ follow by averaging \er{3.3} over the full solid
angle $4 \pi$ at each fixed $D_L$:
\begin{equation}\label{3.4}
<\hspace{-1mm}z\hspace{-1mm}>_{4\pi} = H_4 D_L - \frac 1 2 {H_4}^2 (1 - q_4)
{D_L}^2 + {\cal O}({D_L}^3).
\end{equation}
These equations have the same algebraic form as the corresponding equation in an
FLRW model.

\section{Problems with HS's equation (38)}\label{HSproblems}

\setcounter{equation}{0}

By considering a bundle of null geodesics converging to a point in space, Hirata
and Seljak \cite{HiSe2005} derived an equation relating the quantities in
(\ref{3.4}) to the invariants of flow of a perfect fluid filling the spacetime.
Flanagan's (hereafter F, \cite{Flan2005}) equation is more general, and it
reads, in HS's notation,
\begin{equation}\label{4.1}
{H_4}^2 q_4 = \frac {4 \pi} 3 (\rho + 3p) + \frac 1 3 \left[\dot{u}_{\alpha}
\dot{u}^{\alpha} + \frac 7 5 \sigma_{\alpha \beta} \sigma^{\alpha \beta} -
\omega_{\alpha \beta} \omega^{\alpha \beta} - 2
{\dot{u}^{\alpha}}{}_{;\alpha}\right].
\end{equation}
(HS's result is the subcase $\dot{u}^{\alpha} = 0 = \omega_{\alpha \beta}$.)

Equation (\ref{4.1}) has the appearance of being exact and covariant.
Unfortunately, it is neither. As we show below, its derivations, both by HS and
by Flanagan, contain approximations (some not explicitly spelled out) and
involve averaging over directions, which produces coordinate-dependent results
at locations corresponding to coordinate singularities. Approximations may be
good for some purposes, but not when one intends to show that some quantity is
positive.

Below we note instances where HS and F introduced approximations in their
reasoning.

\subsection{HS's equation (26).}

We have no objections to the reasoning of HS up to their eq. (25). Having
established the relation between the luminosity distance $D_L$ and the redshift
$z$ (their (24)), and between the radiation amplitude and the expansion scalar
$\theta_n$ of the radiation field (their (25)), HS then copy a relation between
$\theta_n$ ($\hat{\theta}$ in HS's notation) and the affine parameter $v$ from
Ref. \cite{BMRi2005} (HS's eq. (26)). This requires a comment.

To obtain this relation one considers another family of rays, originating at the
centre of a source (call it a star for brevity). The star is assumed spherical
and having the radius $R$. One of the rays hits the observer. The values of the
affine parameter along this ray are $v = 0$ at the observer, $v_s$ at the
surface of the star and $(v_s + \Delta v_s)$ at the centre of the star. The
relation is then
\begin{equation}\label{4.2}
\theta_n = \frac 2 {v - (v_s + \Delta v_s)} + {\cal O}[v - (v_s + \Delta v_s)].
\end{equation}
In Ref. \cite{BMRi2005} this is obtained as an approximate solution of the
Raychaudhuri equation for null geodesics, which is
\begin{equation}\label{4.3}
k^{\gamma} \theta_{n, \gamma} + 2 \left({\sigma_n}^2 - {\omega_n}^2\right) +
\frac 1 2 {\theta_n}^2 = - R_{\rho \gamma} k^{\rho} k^{\gamma},
\end{equation}
where $\theta_n = {k^\mu}_{; \mu}$, $\sigma_n$ and $\omega_n$ are, respectively,
the expansion, shear and rotation of the null congruence $k^\mu = \dril
{x^{\mu}} v$ and $R_{\rho \gamma}$ is the Ricci tensor in spacetime.

Equation (\ref{4.2}) without the second term on the right is the exact solution
of (\ref{4.3}) in the case $\sigma_n = \omega_n = 0 = R_{\rho \gamma} k^{\rho}
k^{\gamma}$. Thus the approximation involved in (\ref{4.2}) is that the
contributions to $k^{\gamma} \theta_{,\gamma}$ from shear, rotation and lensing
by matter are negligible compared to ${\theta_n}^2$. These are additional
assumptions about light propagation in spacetime ($\sigma_n$ and $\omega_n$
negligible) and about the spacetime itself ($R_{\rho \gamma} k^{\rho}
k^{\gamma}$ negligible).

However, HS's (38) is used by VFW as if it were exact. When $\sigma_n = 0 =
R_{\rho \gamma} k^{\rho} k^{\gamma}$ hold exactly, the Goldberg -- Sachs theorem
\cite{GoSa1962} applies, which says that the spacetime must be algebraically
special, with $k^\mu$ being the double principal null congruence of the Weyl
tensor. The assumption $R_{\rho \gamma} k^{\rho} k^{\gamma} = 0$, when imposed
on the L--T spacetime, reduces it to the Schwarzschild solution. Therefore, HS's
equation (38), when treated as exact, involves strong assumptions that don't
apply to the L--T model, or even the FLRW model.

\subsection{The two errors that cancelled each other}

The two errors shown below have no meaning for the final result because (in the
approximation in which HS worked) they cancelled out in the end, but they might
be confusing. HS's equation (27) should actually read
\begin{equation}\label{4.4}
\Delta v_s = - \frac R {1 + z},
\end{equation}
i.e. it is the inverse of what they wrote. Similarly, their equation (28) should
actually read
\begin{equation}\label{4.5}
\frac {A(v_s)} {A(0)} = 1 + \frac {v_s} {\Delta v_s}
\end{equation}
(where $A$ is the radiation amplitude), i.e. again the inverse of their result.
Since HS assumed $\left|\Delta v_s\right| \ll \left|v_s\right|$, the above is
approximately equal to
\begin{equation}\label{4.6}
\frac {A(v_s)} {A(0)} = \frac {v_s} {\Delta v_s} = - v_s \frac {1 + z} R,
\end{equation}
which results in HS's eqs. (29) -- (31) being right again. Since we will later
refer to their (31), we copy it here:
\begin{equation}\label{4.7}
z = - K_{ij} n^i n^j {D_L}  - 2 \left(K_{ij} n^i n^j\right)^2 \left({{D_L}
}\right)^2 + \frac 1 2 \ \left(\dot{K}_{ij} + K_{ij|k} n^k + 4 {K_i}^k
K_{kj}\right) n^i n^j \left({ {D_L} }\right)^2 + {\cal O}({D_L}^3).
\end{equation}
The meaning of the symbols in (\ref{4.7}) is as follows. Since HS work under the
assumption that $\omega = 0$ for the cosmic fluid, comoving and synchronous
coordinates exist in which the metric $g_{\alpha \beta}$ has the properties
$g_{00} = 1$, $g_{0i} = 0$, where $x^0 = t$ is the time coordinate and $x^i$, $i
= 1, 2, 3$, are the space coordinates. Then $h_{ij} = - g_{ij}$ is the metric of
a space of constant $t$, and $K_{ij}$ is the second fundamental form of this
space; in these coordinates $K_{ij} = - (1/2) (\pdril {} t) h_{ij} \df - (1/2)
\dot{h}_{ij}$. Then the geodesic null vector $k^{\alpha}$ can be normalised at
the observer so that $k^0 = 1$ and its space components $k^i \df n^i$ form a
3-dimensional unit vector, $h_{ij} n^i n^j = 1$.

\subsection{The problem with averaging}
\label{averprob}

There is a problem with averaging products like $n^i n^j$ and $n^i n^j n^k$ at
coordinate singularities in space, and the centre of symmetry in L--T is a
singularity of the spherical coordinates. For example, a general unit vector in
Euclidean space attached at a point ${\cal O}$, when referred to Cartesian
coordinates and parametrised by spherical angles $(\alpha, \beta)$, has the
components $(n^x, n^y, n^z) = (\sin \alpha \cos \beta, \sin \alpha \sin \beta,
\cos \alpha)$. In this case, the averages\footnote{The averages are defined by
 $$
<\hspace{-1mm}n^i n^j\hspace{-1mm}>_{4\pi} = \frac 1 {4 \pi} \int_0^{\pi} {\rm
d} \alpha \int_0^{2 \pi} {\rm d} \beta n^i n^j \sin \alpha,
 $$
and similarly for $<\hspace{-1mm}n^i n^j n^k\hspace{-1mm}>_{4\pi}$.} are $
\mbox{$<\hspace{-1mm}n^i n^j\hspace{-1mm}>_{4\pi}$} = (1/3) \delta^{ij}$,
$\mbox{$<\hspace{-1mm}n^i n^j n^k\hspace{-1mm}>_{4\pi}$} = 0$,
$\mbox{$<\hspace{-1mm}n^i n^j n^k n^l\hspace{-1mm}>_{4\pi}$} = (1/5)
\delta^{(ij} \delta^{kl)}$. The first and last of these are special cases (in
Euclidean space) of the covariant equations
\begin{eqnarray}
&& \mbox{$<\hspace{-1mm}n^i n^j\hspace{-1mm}>_{4\pi}$} = (1/3) h^{ij},
\label{4.8} \\
&& \mbox{$<\hspace{-1mm}n^i n^j n^k\hspace{-1mm}>_{4\pi}$} = 0, \label{4.9}
\\
&& \mbox{$<\hspace{-1mm}n^i n^j n^k n^l\hspace{-1mm}>_{4\pi}$} = (1/5) h^{(ij}
h^{kl)} \equiv (1/15) \left(h^{ij} h^{kl} + h^{ik} h^{jl} + h^{il}
h^{jk}\right).\label{4.10}
\end{eqnarray}
However, these results of averaging, change when we transform to spherical
coordinates centred at ${\cal O}$. When the components of the unit vector $n^i$
in Euclidean space are referred to the spherical coordinates, they are $(n^1,
n^2, n^3) = (1, 0, 0)$ (with $(x^1, x^2, x^3) = (r, \alpha, \beta)$), and then
the averages are $<~\hspace{-3mm}n^i n^j\hspace{-1mm}>_{4\pi} = \delta^i_1
\delta^j_1$, $<\hspace{-1mm}n^i n^j n^k\hspace{-1mm}>_{4\pi} = \delta^i_1
\delta^j_1 \delta^k_1$. Explicit calculation done in Sec. \ref{anysign} for the
L--T model will show that the general formulae (\ref{4.8}) -- (\ref{4.10}), in
particular (\ref{4.9}), lead to incorrect results when applied at the centre of
symmetry.

Since (\ref{4.1}) is supposed to follow from (\ref{4.7}) by this kind of
averaging, the argument above invalidates (\ref{4.1}) at the centre of symmetry
of the L--T model even as an approximate equation.

\section{The sign of $q_{3'}$ can be any}\label{anysign}
\setcounter{equation}{0}

We will now use (\ref{4.7}) to estimate HS's deceleration parameter $q_{3'}$ at
the centre of the L--T model without averaging over directions. Identifying the
coefficient of $D_L$ in (\ref{4.7}) with the $H_{3'}$ of (\ref{3.3}) we find
from (\ref{3.3}):
\newpage
\begin{equation}\label{5.1} q_{3'} = 1 + \frac 1 {{H_{3'}}^2} \left[- 4
\left(K_{ij} n^i n^j\right)^2 + \left(\dot{K}_{ij} + 4 {K_i}^k K_{kj}\right) n^i
n^j + K_{ij|k} n^i n^j n^k\right],
\end{equation}

In the coordinates of (\ref{2.1}) we have, with $(x^1, x^2, x^3) = (r,
\vartheta, \varphi)$:
\begin{eqnarray}
\left(h_{11}, h_{22}, h_{33}\right) &=& \left(\frac {{R_{,r}}^2} {1 + 2E}, R^2,
R^2 \sin^2 \vartheta\right),\label{5.2} \\
\left(K_{11}, K_{22}, K_{33}\right) &=& \left(- \frac {R_{,r} R_{,tr}} {1 + 2E},
- R R_{,t} - R R_{,t} \sin^2 \vartheta\right), \label{5.3} \\
K_{11|1} &=& \frac {R_{,rr} R_{,tr} - R_{,r} R_{,trr}} {1 + 2E},\label{5.4}
\\
K_{12|2} &=& K_{22|1} = \frac {K_{13|3}} {\sin^2 \vartheta} = \frac {K_{33|1}}
{\sin^2 \vartheta} = R_{,t} R_{,r} - RR_{,tr}, \label{5.5}
\end{eqnarray}
the components not listed are zero.

At a general point of an L--T manifold, the unit spatial vector $n^i$ must obey
$h_{ij} n^i n^j = 1$, and so, in consequence of (\ref{5.2}), its components are
constrained by
\begin{equation}\label{5.6}
n^1 = \pm \frac {\sqrt{1 + 2E}} {R_{,r}}\ \sqrt{1 - R^2 \left[\left(n^2\right)^2
+ \sin^2 \vartheta \left(n^3\right)^2\right]}.
\end{equation}
At the centre of symmetry each vector $n^i$ is necessarily radial, so $n^2 = n^3
= 0$, and (\ref{5.6}) gives
\begin{equation}\label{5.7}
n^1(0) = \frac {\sqrt{1 + 2E}} {R_{,r}},
\end{equation}
which is finite by (\ref{2.9}) and (\ref{a.4}).

However, (\ref{5.6}) and (\ref{5.7}) imply that at the centre of symmetry the
angle-average of (\ref{4.7}) is discontinuous. Namely, as long as the vector
$n^i$ is attached off the centre, its $n^1 = n^r$ component may point in the
direction of increasing $r$ (in which case we take (\ref{5.6}) with the $+$
sign) or toward decreasing $r$ (in which case we take $-$ in (\ref{5.6})). Then
the average comes out zero. But when the vector $n^i$ is attached at the centre
of symmetry, $n^1$ may point only toward increasing $r$, so $n^1 > 0$ on the
whole sphere, while $n^2 = n^3 = 0$. Thus, $(n^1)^3 > 0$ everywhere on the
sphere, while all other components of $n^i n^j n^k$ are zero. Consequently,
$\langle (n^1)^3\rangle_{4\pi} > 0$, while all the remaining averages are zero.
Thus, in the spherical coordinates, $\langle n^i n^j n^k\rangle_{4\pi}$ is
discontinuous at the centre of symmetry.

Let us apply (\ref{5.1}) at the centre of symmetry in the L--T model. Using
(\ref{5.2}) -- (\ref{5.7}) and
\begin{equation}\label{5.8}
H_{3'} = - K_{ij} n^i n^j \llim{r \to 0} \lim_{r \to
0}\left(\frac{R_{,tr}}{R_{,r}}\right)
\end{equation}
in (\ref{5.1}) we find
\begin{equation}\label{5.9}
q_{3'}(0) = \lim_{r \to 0} \left\{- \frac{R_{,r} R_{,ttr}}{R_{,tr}^2} + \sqrt{1
+ 2 E} \left( \frac{R_{,rr}} {R_{,r}  R_{,tr}} - \frac{R_{,trr}}
{{R_{,tr}^2}}\right)\right\}.
\end{equation}

In order to clearly see the contribution of the inhomogeneous part of the
geometry, we will separate $R$ and $E$ into the part that survives in the
Friedmann limit and the part that disappears in that limit. Note that the whole
calculation will be exact, and our definitions for $M$, $E$ and $R$ will ensure
they automatically obey the conditions of regularity at the centre (\ref{2.9}).

We define:
\begin{eqnarray}
R &=& r (S + {\cal P}), \qquad \lim_{r \to 0} {\cal P} = 0, \label{5.10}\\
E &=& r^2 (- k/2 + F), \qquad \lim_{r \to 0} F = 0, \label{5.11}
\end{eqnarray}
where $k$ is a constant (the familiar FLRW curvature index, so it may be of any
sign, also zero) and $S(t)$, $F(r)$ and ${\cal P}(t, r)$ are functions ($S$
being the FLRW scale factor). The terms $r{\cal P}$ and $r^2F$ represent
non-Friedmannian contributions to $R$ and $E$, respectively; with ${\cal P} = F
= 0$ the L--T model reduces to that of Friedmann. In calculating the limits at
$r \to 0$ we will assume that the derivatives of $F$ and $t_B$ do not diverge
too fast, that is
\begin{equation}\label{5.12}
\lim_{r \to 0} \left[r \left(F,_r, {F,_r}^2, F,_{rr}, t_{B,r},
t_{B,rr}\right)\right] = (0, 0, 0, 0, 0).
\end{equation}

Since some of the functions will be complicated, we introduce the following
abbreviations:
\begin{equation}
V \df \sqrt{\frac {2M_0} {S + {\cal P}} - k + 2 F}, \label{5.13}
\end{equation}
\newpage
\begin{equation} {\cal L} \df \frac {2 (S + {\cal P}) F,_r} {k - 2 F} - V
\left[\frac {3 F,_r} {k - 2 F} \left(t - t_B\right) + t_{B,r}\right]
\label{5.14}
\end{equation}
(it can be verified that ${\cal L} = {\cal P},_r$), and the subscript $0$ will
denote the limit at $r \to 0$, thus
\begin{eqnarray}
V_0 &=& \sqrt{\frac {2M_0} {S} - k}, \label{5.15} \\
{\cal L}_0 &=& \frac {2 S F,_r(0)} k - V_0 \left[\frac {3 F,_r(0)} k \left(t -
t_B(0)\right) + t_{B,r}(0)\right]. \label{5.16}
\end{eqnarray}

The formulae for the derivatives of $R$ and their limits at $r \to 0$ are given
in Appendix \ref{Rorig}. We see from there that all the terms entering $q_{3'}$
in (\ref{5.1}) have nonzero and nondivergent limits at $r \to 0$.

For $H_{3'}$ we get from (\ref{5.8}):
\begin{equation}\label{5.17}
\lim_{r \to 0} H_{3'} = \frac {V_0} {S}.
\end{equation}

Using (\ref{5.2}) -- (\ref{5.17}) and (\ref{a.1}) -- (\ref{a.8}) in (\ref{5.9})
we now obtain:
\begin{equation}\label{5.18}
\lim_{r \to 0} q_{3'} = \frac {M_0} {S {V_0}^2} - \frac {2 F,_r(0)} {{V_0}^3} -
\frac {3M_0 / S - k} {S {V_0}^2} \left\{2 t_{B,r}(0) + \frac {2 F,_r(0)} {k V_0}
\left[- 2 S + 3 V_0 \left(t - t_B(0)\right)\right]\right\}.
\end{equation}
The terms proportional to $F_{,r}(0)$, $t_{B,r}(0)$ and $F_{,r}(0)/k$ can each
have any sign, and so each one can make $q_{3'}(0)$ negative. Note that
$F_{,r}(0)$ (which comes from the non-Friedmannian contribution to the energy
function) alone can cause $q_{3'}(0) < 0$ (when $t_{B,r} = 0$), and so can
$t_{B,r}(0)$ alone (when $F = 0$).

At the centre of spherical symmetry, $q_{3'} = q_3 = q_4$. This confirms that
HS's result is not correct at $r = 0$, and a negative $q_4$, as found by INN, is
perfectly possible.

The reason why HS concluded that $q_4 \geq 0$ is that they found $\langle
K_{ij|k} n^i n^j n^k \rangle_{4\pi} = 0$ while averaging (\ref{5.1}) which is
not true at a spherical centre in spherical coordinates. In consequence, they
dismissed all terms in (\ref{5.18}) which come from $K_{ij|k} n^i n^j n^k$, i.e.
all except the first one. With only the first term, one obviously gets $\lim_{r
\to 0} q_{3'} \geq 0$.

Since nothing can be concluded in general about the sign of $q_{3'}(0)$, there
would be no contradiction involved in certain L--T models imitating a negative
FLRW deceleration parameter even if eq. (\ref{5.1}) were exact. But we recall:
this equation being approximate, using it to evaluate the sign of $q_{3'}$
cannot lead to an unambiguous conclusion, so the alleged contradiction was a
nonexistent problem.

There is one more piece of evidence that HS's method of averaging is not
universally correct. Suppose (\ref{5.1}) is applied at the centre of symmetry of
an L--T model. Then each component of the sum in (\ref{5.1}) is invariant under
the group $SO(3)$. Thus, averaging over directions should be an identity
operation and change nothing. This is indeed the case when one uses (\ref{5.3})
-- (\ref{5.5}) for $K_{ij}$ and $K_{ij|k}$, applied at $r = 0$, and (\ref{5.7})
for $n^i(0)$; actually one gets (\ref{5.17}) -- (\ref{5.18}) again. However, if
one uses HS's prescription for averaging, then the whole term $\langle K_{ij|k}
n^i n^j n^k\rangle (0)$ still drops out in spite of being spherically symmetric
and nonzero initially.

\section{Problems with Flanagan's eq. (5)}\label{Fproblems}
\setcounter{equation}{0}

Flanagan's paper \cite{Flan2005} addressed the question of whether superhorizon
perturbations can cause apparent acceleration. He concluded that if their effect
on $q$ is negative, it must be too small to \textsl{be responsible for the
acceleration of the universe.} The paper was not intended to be applied to exact
inhomogeneous cosmological models, so the author should not be blamed for the
incorrect use made later of his equation. But the HS subcase of that approximate
equation was used by VFW as if it were exact.

In this section we follow F's reasoning in order to identify the approximations
assumed along the way.

\subsection{Comments to F's reasoning up to eq. (13)}

The correct form of Flanagan's eq. (9) is:
\begin{equation}\label{6.1}
\overline{u}^{\alpha}(x, x') = u^{\alpha}(x) + u^{\alpha \beta}(x) \sigma_{;
\beta} (x, x') + \frac 1 2 u^{\alpha \beta \gamma}(x) \sigma_{; \beta} (x, x')
\sigma_{; \gamma} (x, x') + O(s^3).
\end{equation}
The meaning of the symbols is as follows: $x$ is the set of coordinates of the
observer at the point $P$, $x'$ is the set of coordinates of the light source at
the point $Q$, $u^{\alpha}(x)$ is the four-velocity of the cosmic medium at $P$
(the observer is comoving), $\overline{u}^{\alpha}(x, x')$ is the four-velocity
of the medium parallel-transported from $Q$ to $P$ along the light ray,
$\sigma(x, x')$ is Synge's \cite{Syng1960} world function for $P$ and $Q$, and
the coefficients are $u^{\alpha \beta}(x) = - u^{\alpha; \beta}(x)$, $u^{\alpha
\beta \gamma}(x) = u^{\alpha; (\beta \gamma)}(x)$.

The quantity in (\ref{6.1}) is apparently only needed to calculate the redshift,
which F writes as
\begin{equation}\label{6.2}
1 + z = \frac {\overline{u}^{\alpha} k_{\alpha}} {u^{\alpha} k_{\alpha}},
\end{equation}
where $k_\alpha$ is the tangent vector to the ray at $P$. Usually the numerator
is taken simply at $Q$, without parallel-transporting it to $P$ \cite{Elli1971}.
Nevertheless, it is easy to verify that (\ref{6.2}) is equal to the usual
formula.

F then considers two families of light rays: one converging at $P$, with the
affine parameter $s$ and tangent vectors $k$ normalised so that $k_{\alpha}
u^{\alpha} = -1$ at $P$, and another one diverging from $Q$, with the affine
parameter $\overline{\lambda}$ and tangent vectors $l$ normalised so that
$l_{\alpha} u^{\alpha} = -1$ at $Q$. In consequence of this, and of (\ref{6.2}),
the two affine parameters are related by
\begin{equation}\label{6.3}
\overline{\lambda} = s (1 + z).
\end{equation}

\subsection{The $\theta_n$ approximation}

Below his (13), F writes: \textsl{We choose the normalization of $A$ so that $A
\approx 1/\overline{\lambda}$ for $\overline{\lambda} \to 0$ near $Q$.} This is
an assumption equivalent to HS's (26) -- our (\ref{4.2}) ($A$ is the same
quantity as in (\ref{4.5})), and F makes use of it further on (see below our eq.
(\ref{6.7})). This \textsl{normalization} also defines the units for
$\overline{\lambda}$ as the units of distance.

\subsection{F's equation (14)}

In the paragraph of F's paper that contains eqs. (14) -- (16) the quantities
referring to $P$ and to $Q$ are mixed up. The text below is our interpretation
of this segment of F's reasoning, in which we take care about properly
distinguishing between these two points.

{}From the definition of the energy flux at $P$ one obtains:
\begin{equation}\label{6.4}
\dr {E_P} {t \, d^2 A} = T_{\alpha \beta} (P) u_P^{\alpha} u_P^{\beta} = {A_P}^2
\left(l_{\alpha} u^{\alpha}\right)^2_P = {A_P}^2 \left(k_{\alpha}
u^{\alpha}\right)^2_P / (1 + z)^2 = {A_P}^2 / (1 + z)^2,
\end{equation}
whereby a rewriting error was corrected above: F's $\left(k_{\alpha}
u^{\alpha}\right)^2$ should read $\left(l_{\alpha} u^{\alpha}\right)^2$. The
following information was fed into (\ref{6.4}) at the consecutive equality
signs: the definition of energy flux, the definition of the radiation
energy-momentum tensor at $Q$, equation (\ref{6.2}), the normalization of
$k^{\alpha}$ at $P$.

Integrating the energy flux {\em at $Q$} over a sphere of small radius
$\overline{\lambda}_Q$, one obtains for the luminosity at $Q$, again from the
definition:
\begin{equation}\label{6.5}
\dr E t = {A_Q}^2 \left(l_{\alpha} u^{\alpha}\right)^2_Q \times 4\pi
{\overline{\lambda}_Q}^2 = 4 \pi {A_Q}^2 {\overline{\lambda}_Q}^2;
\end{equation}
the simplification occurs because of the normalization assumed at $Q$. Putting
(\ref{6.2}) and (\ref{6.3}) into the definition of the luminosity distance
(which we denote here by $D_L$ for consistency with the previous sections) one
obtains
\begin{equation}\label{6.6}
A_Q \overline{\lambda}_Q = \frac {D_L A_P} {1 + z}.
\end{equation}

In the next step F uses the assumed normalization of $A_Q$ to write the above as
$D_L = (1 + z) / A$, but then gives up on it and uses (\ref{6.3}) to rewrite
(\ref{6.6}) as follows:
\begin{equation}\label{6.7}
D_L = (1 + z)^2 s_Q A_Q / A_P = (1 + z)^2 s_Q / (A_P \overline{\lambda}_Q)
\end{equation}
(the second equality follows by the assumed normalization of $A$ at $Q$). Then F
cites Visser \cite{Viss1993} for the result
\begin{equation}\label{6.8}
A_P \overline{\lambda}_Q = 1 + O({s_Q}^2).
\end{equation}
Thus finally, up to $s^2$-terms:
\begin{equation}\label{6.9}
D_L \approx (1 + z)^2 s_Q,
\end{equation}
which is the same equation as HS's (29). Just as in the HS paper, this equation
is approximate, where in addition to HS's (26) (our (\ref{4.2})), Flanagan used
one more approximation -- our (\ref{6.8}).

{}From (\ref{6.9}) and from eqs. (1) and (11) in F's paper, F's equations (15)
and (16) follow by simple substitution.

\subsection{Choice of averaging procedure}
\label{cap}

Below his eq. (16), Flanagan notes that there is no unique way to choose\ \ the
order\  in which\ \  the coefficients in
\newpage
\noindent the $D_L(z)$ equation should be averaged over directions, but
dismisses this problem by saying:

\medskip

\noindent \textsl{Thus the different averaging prescriptions give different
answers. However the fractional differences are of order $\sigma^2/\theta^2$,
which we have argued above is of order $\varepsilon$ and is small.}

\medskip

\noindent This is correct in the context of his paper. However, this indicates
that there is a further approximation involved in the reasoning, and this one is
quite arbitrary and unpredictable. Namely, it depends on the will of the person
doing the calculation exactly which quantity he or she wishes to average first.
F himself uses $H_0 = \langle A^{-1} \rangle$ to obtain the $H_0$ of his eq.
(4), but $H_0^{-1} = \langle A \rangle$ for computing his $J$. This is perhaps
the strongest indication that it makes no sense to use the resulting final
equation for determining the sign of any quantity. Note that the same problem
exists for the HS derivation, but it was not mentioned in the HS paper.

\subsection{The problem with rotation}

The averaging over directions is defined only in a certain 3-space $S_3$, not in
spacetime. Without fixing $S_3$, the directional angles of the light rays are
not defined. When the rotation of the cosmic fluid is zero, as in HS's paper,
$S_3$ is the 3-space orthogonal to the fluid's flow lines.

However, when the cosmic fluid has nonzero rotation, such $S_3$ do not exist --
not even locally, because the 3-volume elements locally orthogonal to the flow
lines, when followed around any flow-tube, refuse to connect up to a 3-space.
One could consider a 3-space of constant time-coordinate, which is not
orthogonal to the flow lines, but then the metric in this space is not the
$h_{\alpha \beta} = g_{\alpha \beta} + u_{\alpha} u_{\beta}$ assumed by F. Thus,
in the presence of rotation, the whole calculation should be reformulated.
Without that, for the averaging over angles considered by F there exists no
space in which it occurs.

\section{VFW's \textsl{weak singularity} is
not a singularity} \setcounter{equation}{0}

In the paragraph containing their eq. (2.18) VFW say:

\textsl{We expand the density (2.5) to second order in $r$ as
 $$
\rho(r, t) = \rho_0(t) + \rho_1(t)r + \rho_2(t)r^2 + {\cal O} (r^3). \eqno(2.18)
 $$
The weak singularity occurs when $\rho_1(t)$ is nonzero, in which case the
gravitational field is singular since $\square {\cal R} \to \pm \infty$ as $r
\to 0$, where ${\cal R}$ is the Ricci scalar. In other words, second derivatives
of the density diverge at the origin, independent of where observers may be
located. This is true both in flat spacetime and in the curved LTB metric when
we have a density profile of the form (2.18). The singularity is weak according
to the classification scheme of the literature on general relativity [28].}
(Ref. [28] in this quotation is \cite{Tipl1977}.)

In truth, whether a curvature singularity is there is decided by curvature
alone, and not by secondary constructs like the d'Alembertian of the curvature
scalar. We do not know about any physical or geometrical interpretation of this
quantity, and VFW do not mention any, nor have they shown that their `weak
singularity' causes any problems at the origin. As we show below the curvature
of the L--T model at the centre of symmetry is nonsingular provided that $\rho$
is finite there.

For the metric (\ref{2.1}), the orthonormal tetrad components of the curvature
tensor, in the basis defined by $e^0 \df {\rm d} t$, $e^1 \df R_{,r} {\rm d} r /
\sqrt{1 + 2E}$, $e^2 \df R {\rm d} \vartheta$, $e^3 \df R \sin \vartheta {\rm d}
\varphi$, are
\begin{eqnarray}\label{7.1}
R_{0 1 0 1} &=& \frac {2M} {R^3} - \frac {M_{,r}} {R^2R_{,r}}, \qquad R_{0 2 0
2} = R_{0 3 0 3} = - \frac M {R^3}, \nonumber \\
R_{1 2 1 2} &=& R_{1 3 1 3} = \frac M {R^3} - \frac {M_{,r}} {R^2R_{,r}}, \qquad
R_{2 3 2 3} = - \frac {2M} {R^3} ~.
\end{eqnarray}
The quantities given above are all scalars, so any scalar polynomial in the
curvature components will be a polynomial in the quantities given in
(\ref{7.1}). If these are nonsingular, then there will be no scalar polynomial
curvature singularity.

Let $r = r_c$ be the radial coordinate corresponding to the centre of symmetry
$R = 0$. The requirement of no point-mass at the centre implies $M(r_c) = 0$. It
can be seen that with $M(r_c) = R(t, r_c) = 0$ all we need to make the curvature
nonsingular at $r = r_c$ is a finite limit of $M/R^3$ at $r = r_c$. We have,
using (\ref{2.4})
\begin{equation}\label{7.2}
\lim_{r \to r_c} \frac M {R^3} = \lim_{r \to r_c} \frac {M_{,r}} {3 R^2 R_{,r}}
= (4 / 3) \pi G \rho(t, r_c),
\end{equation}
so the curvature is finite at $r = r_c$ if and only if $\rho(t, r)$ is.

As regards geodesic completeness, the geodesic equation is well behaved through
the origin. Geodesics passing through the origin must be purely radial, and
radial geodesics stay radial. The only Christoffel symbols that diverge at $R =
0$ are $\Gamma^\theta{}_{r\theta} = \Gamma^\phi{}_{r\phi} = R'/R$, but these do
not appear in the equations of radial geodesics, so cannot cause any problem.

The condition $\square {\cal R} \to \pm \infty$ is not a criterion for a
singularity in any accepted sense. The paper of Tipler \cite{Tipl1977} contains
a proposal of a definition of a {\em strong singularity}, but does not mention a
`weak singularity' anywhere and contains no `classification scheme'. For a
singularity to be `weak' (i.e. not strong), it first has to be a singularity. An
accurate description of the feature that VFW had in mind is `the density profile
is not $C^1$ through the origin'.

Therefore, VFW's \textsl{weak singularity} is not a singularity, so its presence
is no reason to dismiss models containing it. There is nothing wrong with a jump
in the density gradient at the centre. Since our Galaxy is centrally
concentrated and has a central black hole, a pointed density profile is a good
approximation. The NFW density profile \cite{NFWh1996} for galaxy clusters is
divergent at the centre. A jump in the density and its gradient exists, for
example, at the surface of the Earth, thus we deal with a `weak singularity' in
everyday life.

\section{Transcritical solutions are generic}
\label{TrCrGen}

\setcounter{equation}{0}

A large part of VFW's paper is devoted to considering the \textsl{inverse
problem} and arguing that only FLRW models have rays that cross the apparent
horizon --- what they call the \textsl{critical point}.  We here dispel the
impression they convey, that \textsl{transcritical} solutions are well-nigh
impossible to find for non-homogeneous models that account for observations. We
emphasise (a) that all L--T models with a decelerating expansion phase have
apparent horizons that are crossed by large families of light rays, and (b) no
cosmological model has been shown to solve the inverse problem (as distinct from
the fitting problem) relative to real data.

VFW correctly point out that the differential equations (DEs) of the
\textsl{inverse problem}, i.e. the DEs determining the L--T model that
reproduces given observational data, contain terms that would diverge unless a
certain condition is met. They call such loci \textsl{critical points}, and they
correctly claim that such points are generic to these DEs. This was actually
pointed out by Mustapha, Hellaby and Ellis \cite{MHEl1997}, solved by Lu and
Hellaby \cite{LuHe2007}, discussed as an observational feature in
\cite{Hell2006}, and used to advantage by McClure and Hellaby \cite{McHe2008}.
These papers develop a numerical procedure for extracting the metric of the
cosmos from observational data.\footnote{This problem persists even in the
fluid-ray coordinates (observer coordinates): in \cite{HelAlf09}, the
differential equations (59)-(65) contain $\pdil{C}{z}$ in the denominator, and
on the observer's past null cone $w = w_0$, this is $\pdil{\Rh}{z}$ here, so it
is again zero at the apparent horizon.}

In their analysis of the inverse problem, VFW consider only flat L--T models,
which are a subset of measure zero in the family of all possible L--T models.
They say that FLRW models provide examples of \textsl{transcritical} solutions
(i.e. those that do not cause any divergence of the DEs), but they fail to find
any other viable solution.\footnote{An L--T model is fully specified by two
physical functions. We agree that a random combination such as $E = 0$ and an
arbitrarily chosen $D_L(z)$, specified via $r_{FRW}(z)$ or $V(z)$, may well
produce an L--T model with unrealistic features.} They then conclude that
physically reasonable solutions {\em must be very exceptional indeed}, and
although they have not tried the $E \neq 0$ case, they opine that {\em generic
solutions with $E(r) \neq 0$} would have all the \textsl{singularities} and
\textsl{observational problems} they encountered, and in this case too
transcritical solutions do {\em not appear to be likely.}

In assessing the likelihood of finding a `transcritical solution' it is
important to understand the physics and geometry of the problem. For any {\em
given} cosmological metric, the DEs that determine the path of the light ray and
the variation of $z$ and $D_A(z)$ along it (for example \er{2.5}, \er{8.2},
\er{2.6}) are free of critical points. However, the path followed by an incoming
light ray depends on the geometry it passes through, and this affects the
observations, $z$, $D_A(z)$ etc. All known expanding cosmologies with a
deceleration phase have a past apparent horizon (AH$^-$); relative to a given
worldline (observer), it is the locus where incoming light rays overcome the
cosmic expansion and start to make progress inwards. The maximum in the area
distance $D_{Am} = \Rh_m$ occurs where an observer's PNC crosses her past
apparent horizon. The fact that $\dril {\Rh} z = 0$ at this maximum is the cause
of the critical behaviour. This behaviour is a generic feature of the L--T and
Friedmann models. We express this in the following theorems. We consider only
models that: (i) have a big bang and after 14 billion years are still expanding,
so any recollapse is to the future, and times well beyond the present are not
considered; (ii) have no shell crossings \cite{HeLa1985}; (iii) are large, e.g.
if the curvature is positive, then any spatial maximum where $M_{,r} = 0 =
R_{,r}$ is well beyond observational range, which means $M_{,r} > 0$ and $R_{,r}
> 0$; (iv) if $\Lambda > 0$ its effect is only discernable in fairly distant
observations. We will call such a model an rLTc: a realistic L--T cosmology.

\newtheorem{Thm}{Theorem}

\begin{Thm}
In every rLTc, every light ray arriving at the origin is a transcritical ray.
\end{Thm}

\noindent {\bf Proof}~~ The areal radius $R(t,r)$ obeys
 \begin{align}
   R_{,t} & = \ell \sqrt{\frac{2 M}{R} + 2 E + \frac{\Lambda R^2}{3}}\;
   \label{8.1}
 \end{align}
where $\ell = +1$ if $R$ is increasing (expansion) and $-1$ if it is decreasing
(contraction). Worldlines with $2 E \geq - (9 M^2 \Lambda)^{1/3}$ are
ever-expanding and the rest are recollapsing. Those that recollapse reach their
maximum $R$ where $6 M + 6 E R + \Lambda R^3 = 0$ and this maximum is $\leq (3
M/\Lambda)^{1/3}$.

The incoming radial light rays satisfy \er{2.5}, so the variation of $R$ along a
ray is
 \begin{align}
   \dr{\Rh}{r} & = \left[ R_{,t} \dr{\th}{r} + R_{,r} \right]_\wedge
      = \left( - \ell \frac{\sqrt{2 M / \Rh + 2 E
      + \Lambda \Rh^2 / 3}\;} {\sqrt{1 + 2 E}\;} + 1 \right)
      \widehat{R_{,r}}
      = \frac{- \widehat{R_{,r}}}{\sqrt{1 + 2 E}\;} \td{\Rh}{t}
   \label{8.2}
 \end{align}
The past apparent horizon is the locus where $\dril{\Rh}{r} = 0$ for all rays,
which implies
 \begin{align}
   6 M = 3 R - \Lambda R^3  \label{8.3}
 \end{align}
and the past apparent horizon has $\ell = +1$. For a given $M > 0$, \er{8.3} has
either two, one or zero solutions in the range $R > 0$. The smaller $R$ solution
corresponds to the regular ($\Lambda = 0$) AH, and the larger one corresponds to
the de Sitter horizon, but the two merge on the $3 M \sqrt{\Lambda}\; = 1$
worldline.

Near the origin, $R(t, 0) = 0~ \forall~t$, regularity
\cite{CHPhD85,HuMaMa98,MuHe2001,PlKr2006} requires $R \sim M^{1/3}$ and $2 E
\sim M^{2/3}$, so by \er{8.1} $R_{,t} \to 0$ and by \er{8.2} $\dril{\Rh}{r} \to
R_{,r}$ there.  This means $6M < 3 R - \Lambda R^3$ for incoming rays near the
origin.  Following the rays outwards and back in time, \er{2.5} shows $r$ and
therefore $M$ is increasing, and \er{8.2} shows $\Rh$ is increasing near the
origin.  Provided $R_{,r}$ stays positive as assumed, \er{2.5} may be integrated
all the way to the big bang; and here we have $M > 0$ and $R \to 0$, so $6M
> 3
R - \Lambda R^3$. Therefore the rays have crossed AH$^-$.  This argument applies
to every ray arriving at the origin before the big crunch, if there is a crunch.

The DEs for the general L--T `inverse problem' are given in
\cite{LuHe2007,McHe2008}, and it is evident that the critical points are where
$\dril{\Rh}{r} = 0 = \dril{\Rh}{z}$.  If a ray crosses the past AH, then by
definition it (a) has $\dril{\Rh}{r} = 0$ there, and therefore (b) it is
transcritical.
 ~$\Box$

If $\Lambda > 0$ then there could%
\footnote{In $E < 0$ models, it is possible the maximum $M$ does not reach $1/(3
\sqrt{\Lambda}\;)$.} be worldlines with $M > 1/(3 \sqrt{\Lambda}\;)$. These
worldlines never encounter a solution to \er{8.3}, so every incoming ray
crossing them has $\dril{\Rh}{r} < 0$ and $\dril{\Rh}{t} > 0$. In ever-expanding
models, AH$^-$ asymptotically approaches $R = \sqrt{3/\Lambda}$ (and $M = 0$)
towards the future, and it separates incoming light rays that always have
increasing $R$ from those that reach the origin. Therefore there are rays
emerging from the bang that neither cross AH$^-$, nor reach the origin.

\begin{Thm}
Every incoming radial light ray in every $\Lambda = 0$ L--T model with a big
bang is a transcritical ray.
\end{Thm}

\noindent {\bf Proof}~~~ When $\Lambda = 0$, \er{8.3} becomes the familiar
 \begin{align}
   R = 2 M
 \end{align}
and such apparent horizons have been well studied in \cite{Hell1987,KrHe2004b}.
In this case, every worldline encounters AH$^-$, which is at $R > 0$ except at
an origin, and every light ray starts on the big bang. Therefore, of the rays
emerging from the bang and arriving at the crunch or $R = \infty$, only those
starting at an origin may be said to have not crossed the past AH. Models
usually have one origin, but closed models normally have two. The rays emerging
from the local origin are outgoing, not incoming, and the rays emerging from an
antipodal origin may eventually be locally incoming, but in an rLTc this would
happen long after the present, if at all.
 ~$\Box$

\begin{Thm}
Transcritical rays of inhomogeneous models that reproduce the observed $D_A(z)$
do exist.
\end{Thm}

\noindent {\bf Proof}~~~ See e.g. \cite{AAGr2006,Hell2009}. The point here is
that if a chosen model produces a $D_A(z)$ and $D_L(z)$ that fits the
observations for a given PNC, out past a maximum in $D_A$, then that PNC is
necessarily transcritical.
 ~$\Box$

\begin{Thm}
Any reasonable $D_A(z)$ can be reproduced by a $\Lambda = 0$ L--T model.
\end{Thm}

\noindent {\bf Proof}~~~ This is a special case of the theorem in
\cite{MHEl1997}, which claims any pair of reasonable observational functions,
$D_A(z)$ or $D_L(z)$ and $\mu n(z)$ can be reproduced by an L--T model. Here
$n(z)$ is the number density of sources (e.g. galaxies) in redshift space, and
$\mu$ is the mean mass per source, which may also vary with $z$. Essentially the
2 arbitrary functions in an L--T model allow one to reproduce 2 observational
relations. In fact \cite{Hell2006,LuHe2007,McHe2008} have demonstrated how to
solve the `inverse problem' numerically for both observational functions, and
shown that the conditions at the AH do impose a significant constraint that
holds only at a single $z$ value.  But with only $D_L(z)$ given, the 2 L--T
functions allow the observations and the constraint to be fitted easily.
 ~$\Box$

Together these results show that there is a plentiful supply of transcritical
solutions amongst the L--T models, and that there is a well defined procedure
for extracting an L--T model that fits the observed $D_A(z)$.

More important, however, is the fact that VFW do not try to find an L--T model
that reproduces real observational data, but the one that reproduces a chosen
idealised function for $D_L(z)$.

VFW write as if FLRW models are known to solve the \textsl{inverse problem}. To
date the `inverse problem' has never even been attempted with real data. What is
usually done is to solve the fitting problem: choose your favorite type of
cosmological metric and find the free functions and parameters that make it best
fit the data.

Even models that are arbitrarily close to reproducing the observations will fail
this test. For example, take the `observational' relations of a given FLRW
model, add small random or systematic errors, and use this `data' as input to
the inverse problem. The DEs of the inverse problem will return the given FLRW
model quite closely, but \er{8.3} will not be {\it exactly} satisfied where
$\dril{\Rh}z = 0$, so they will still diverge (see \cite{McHe2008}). The chances
of an FLRW light ray being exactly transcritical relative to real observational
data are distinctly less than for an L--T model.

The failure of a proposed solution to be transcritical does not indicate any
physical problem with the underlying L--T or FLRW model; at worst it indicates
that a particular ray in a particular model does not (exactly) reproduce the
given observational function(s).

As discussed in \cite{Hell2006,LuHe2007,McHe2008,Hell2009,ArSt2007}, when
attempting to reproduce {\it two} observational functions with an L--T model,
the AH constraints {\it do} provide a significant challenge. In overcoming this
challenge it was shown \cite{Hell2006,McHe2008} the AH can give us useful
observational information.

\section{Conclusions}

We have corrected a number of inaccurate statements appearing in the HS
\cite{HiSe2005}, Flanagan \cite{Flan2005} and VFW \cite{VFWa2006} papers. At
least two of these misunderstandings are widespread and therefore it is worth
putting the record straight.

The first error-correction to be made is that a conical density profile at the
origin is not a singularity in any accepted sense, and there are no physical
problems associated with it. In some situations a conical profile may be a good
model of structures that are strongly centrally concentrated.

Secondly, critical points in the inverse problem are generic. When solving the
`inverse problem' as defined in Sec.~\ref{TrCrGen} for a realistic L--T or FLRW
model, and finding that the differential equations diverge at some redshift, one
must be aware that this is an inevitable consequence of the apparent horizon,
where the diameter distance is maximum. The chances that real observational data
{\em exactly} satisfy the full transcriticality conditions for any given model,
homogeneous or not, are essentially zero. Such divergent behaviour does not
indicate any physical problem with the underlying L--T or FLRW model, and
transcritical solutions in a variety of models that closely reproduce the
observed $D_L(z)$ do exist. Actually this point seems not to be well understood
and other examples of this misunderstanding can be found in the literature (see,
e.g. \cite{INNa2002,ChRo2006}).

Our point three is that one should be careful when extending FLRW definitions of
some functions to other cosmological models.  In most such cases, when
inhomogeneities are present, the FLRW parameters do not mean the same thing as
in the homogeneous case.

It is very common to get acceleration and supernova dimming mixed up.  The
interpretation of the supernova dimming as an accelerated expansion (which is
due to the use of FLRW models) has firmly taken root, which is why some have
focussed on acceleration in the fluid expansion. In the L--T models, the
`acceleration' defined from $z(D_L)$ using (\ref{3.3}) exhibits a behaviour
different from the $q_1$ of the fluid flow defined by (\ref{2.1}), even though
they coincide for FLRW models. We hope our paper will help establish the
difference between `accelerated expansion' and `supernova dimming' in the
observed magnitude--redshift relation.

We summarise and discuss different misunderstandings related to inhomogeneous
cosmology in Table \ref{tab1}.

\begin{table*}
\caption{Summary of misconceptions and misunderstandings related to
inhomogeneous cosmology. } \label{tab1}
\begin{tabular}{ll}  \hline
\hspace{0.8cm} Misconceptions and misunderstandings & \hspace{0.8cm} Corrections
and clarifications \hspace{1cm}  \\ \hline
\multicolumn{2}{c}{Weak singularity} \\
 & \\
\pb{75mm}{
$\Box {\cal R} \to \infty$ is a singularity. \\
Models with  $\Box {\cal R} = \infty$ are unphysical. } & \pb{95mm}{ $\Box {\cal
R} \to \infty$ is not a curvature singularity and has no physical
interpretation. Many objects have $\Box {\cal R} = \infty$, among them the Earth
at its surface.} \\
& \\
\hline
\multicolumn{2}{c}{Deceleration parameter} \\
 & \\
 \pb{75mm}{
There are general theorems that prohibit $q_0 < 0$ where $q_0 \equiv q(z=0)$
presented in Refs. \cite{Flan2005,HiSe2005}. } & \pb{95mm}{ There are two
distinctive deceleration parameters: $q^{obs}$ based on a Taylor expansion of
the luminosity distance and the invariantly defined $q^{inv}$ which measures the
acceleration of expansion. If $\Lambda=0=\omega_{ab}=\dot{u^\alpha}$ and
$\rho+3p>0$ then $q^{inv}>0$, however for the same case $q^{obs}$ may be
negative. F's \& HS's equations relating $q^{obs}$ to the flow invariants are
approximate and coordinate-dependent, thus they do not exist as covariant laws.
The approximately correct intermediate relation does not exclude $q^{obs}<0$  in
the L--T model.
}\\
& \\
\pb{75mm}{ There is a local singularity in models with $q_0 < 0$. } & \pb{95mm}{

$\Box {\cal R} \to \infty$ when $q_0<0$ and the observer is at the centre of
spherical symmetry. However, this is not a singularity. Away from the the centre
of spherical symmetry $q_0<0$ does not imply divergence of $\Box {\cal R}$.
} \\
& \\
\pb{75mm}{ Other singularities arise in models with $q(z)<0$. In such models
there is a class in which $q(z) = -1 + \frac{1+z}{H(z)} \frac{dH(z)}{dz} \to
\infty$ at some $z$, where $H(z) = \left[ \frac{d}{dz} \left( \frac{D_L}{1+z}
\right) \right]^{-1}$. } & \pb{95mm}{ $q$ defined in this way diverges when
$\frac{d}{dz} \left( \frac{D_L}{1+z} \right) = 0$ but this is not a singularity.
In fact, if one applies  these definitions to a zero-$\Lambda$ dust FLRW model,
one obtains  regions where
$q< 0$ when $\Omega_{k0} > 0.6$. \\
This is because such $H(z)$ only applies when $k = 0$.
} \\
& \\ \hline
\multicolumn{2}{c}{Inverse problem} \\
& \\
\pb{75mm}{ In solving for the model that gives a selected $D_L(z)$, one
encounters a `pathology' or `critical point', beyond $z\sim 1$ and the solution
generally breaks down there. } & \pb{95mm}{ The `critical point' is the apparent
horizon, where the diameter distance is maximum. It is a general property of
expanding decelerating $\Lambda = 0$ models such as L--T and FLRW, long known in
the FLRW case. This point requires special treatment, but has useful properties
\cite{Hell2006,LuHe2007,McHe2008}.
} \\
& \\
\pb{75mm}{ Only FLRW models have ``transcritical" solutions. } & \pb{95mm}{
``Transcritical" rays, that cross the apparent horizon, are generic in  L-T and
FLRW models. With small errors in observational data, all L-T and FLRW models
will technically fail the apparent horizon conditions, but this can be fixed
\cite{McHe2008}. Arbitrary choices such as $E = 0$ and a given $D_L(z)$ may well
fail to be transcritical.
} \\
& \\
\hline
\multicolumn{2}{c}{Fitting observations} \\
& \\
\pb{75mm}{ Inhomogeneous models are exposed to singularities. Non-singular
models are exceptional and rigid, hence unable to account for observations. } &
\pb{95mm}{ Inhomogeneous models like the L--T models include the FLRW models as
a subcase. Thus, if the FLRW models are considered good enough for cosmology,
then the L--T models can only be better: they constitute an {\em exact
perturbation} of the FLRW background, and can reproduce the latter as a limit
with an arbitrary precision.
} \\
& \\
\pb{75mm}{ $q^{obs}_0 <0 $ is essential to account for observations. } &
\pb{95mm}{ Derivation of $q^{obs}$ in an inhomogeneous model involves
approximations -- one of them linearity, so the sign of $q^{obs}$ cannot be
determined. One can have a very good fit to observations with $q^{obs}_0 >0$
\cite{BW09,YKN08}.
} \\
&  \\
\hline
\end{tabular}
\end{table*}

\bigskip

\noindent {\bf Acknowledgements}

\medskip

MNC wants to thank the Department of Mathematics and Applied Mathematics of the
University of Cape Town for hospitality while part of this work was carried out,
and particularly Charles Hellaby and Peter Dunsby for their nice welcome. The
authors are grateful to Kayll Lake for his interest in their work and useful
comments.

\appendix

\section{Behaviour of derivatives of $R(t, r)$ in the neighbourhood of the
symmetry centre}\label{Rorig}

\setcounter{equation}{0}

The formulae below are obtained using the coordinate $r$ introduced in
(\ref{2.8}), the symbols introduced in (\ref{5.10}) -- (\ref{5.11}), and eq.
(\ref{2.2}). We show the intermediate expressions in order to demonstrate that
all terms that could cause divergencies cancel out before the limit is taken.
The first two equations follow trivially from (\ref{2.2}).
\begin{eqnarray}
R_{,t} &=& \sqrt{\frac {2M} R + 2E} = r V \llim{r \to 0} 0, \label{a.1}
\end{eqnarray}
 \newpage
\begin{eqnarray}
R_{,tt} &=& - \frac M {R^2} = - \frac {M_0 r} {(S + {\cal P})^2} \llim{r \to 0}
0, \label{a.2}
\end{eqnarray}
To evaluate the behaviour of $R_{,r}$ and $R_{,rr}$ in the neighbourhood of $r =
0$ we use the following equation \cite{HeLa1984}:
\begin{equation}\label{a.3}
R_{,r} = \left(\frac {M_{,r}} M - \frac {E_{,r}} E\right)R + \left[\left(\frac 3
2 \frac {E_{,r}} E - \frac {M_{,r}} M\right) \left(t - t_B\right) -
t_{B,r}\right] R_{,t}.
\end{equation}
The above applies also with $E = 0$ if we neglect the $E_{,r}/E$ terms (as
verified by direct calculation). {}From (\ref{a.3}) we obtain:
\begin{equation}\label{a.4}
R_{,r} = S + {\cal P} + r {\cal L} \llim{r \to 0} S.
\end{equation}
The next equations below follow by consecutively differentiating (\ref{a.1}) --
(\ref{a.3}) by $r$ and using (\ref{a.3}) -- (\ref{a.4}) in the result:
 \begin{equation}
R_{,ttr} = - \frac {M_{,r}} {R^2} + \frac {2 MR_{,r}} {R^3} = - \frac {M_0} {(S
+ {\cal P})^2} + \frac {2 M_0 r {\cal L}} {(S + {\cal P})^3} \llim{r \to 0} -
\frac {M_0} {S^2}, \label{a.5}
\end{equation}
\begin{equation}
R_{,tr} = \frac 1 {R_{,t}} \left(\frac {M_{,r}} R - \frac {MR_{,r}} {R^2} +
E_{,r}\right) = V + \frac r V\ \left[F_{,r} - \frac {M_0 {\cal L}} {(S + {\cal
P})^2}\right] \llim{r \to 0} V_0, \label{a.6}
\end{equation}
\begin{eqnarray}
R_{,trr} &=& - \frac 1 {{R_{,t}}^3} \left(\frac {M_{,r}} R - \frac {MR_{,r}}
{R^2} + E_{,r}\right)^2 \begin{array}{l} \vspace{8mm} \end{array} + \frac 1
{R_{,t}} \left(\frac {M_{,rr}} R - \frac {2 M_{,r} R_{,r}} {R^2} - \frac {M
R_{,rr}} {R^2} + \frac {2 M {R_{,r}}^2} {R^3} + E_{,rr}\right)
 \nonumber \vspace{2mm} \\
&=& \frac 1 V\ \left\{- \frac {M_0} {(S + {\cal P})^2} R_{,rr} + 2 F_{,r} + r
\left[F_{,rr} + \frac {2 M_0 {\cal L}^2} {(S + {\cal P})^3}\right]\right\} -
\frac r {V^3}\ \left[F_{,r} - \frac {M_0 {\cal L}} {(S + {\cal P})^2}\right]^2
\nonumber \\
&& \llim{r \to 0} \frac 1 {V_0} \left[- \frac {M_0} {S^2} R_{,rr}(0) + 2
F_{,r}(0)\right],\label{a.7}
\end{eqnarray}
\begin{eqnarray}\label{a.8}
R_{,rr} &=& \frac 1 {- k/2 + F}\ \left[- (S + {\cal P}) + \frac 3 2\ V \left(t -
t_B\right)\right] \left[2 F_{,r} + r \left(F_{,rr} - 2 \frac {{F_{,r}}^2} {- k/2
+ F}\right)\right] \nonumber \\
&-& V \left[2 t_{B,r} + r \left(t_{B,rr} + \frac {F_{,r} \, t_{B,r}} {2 (- k/2 +
F)}\right)\right] \nonumber \\
&+& \frac r V\  \left[\frac 3 2\ \frac {F_{,r}} {- k/2 + F} \left(t - t_B\right)
- t_{B,r}\right] \left[F_{,r} - \frac {M_0 {\cal L}} {(S + {\cal P})^2}\right]
\nonumber \\
\llim{r \to 0} &\ &  - \frac {2 F_{,r}(0)} k \left[- 2 S + 3 \left(t -
t_B(0)\right) V_0\right] - 2 t_{B,r}(0) V_0. \nonumber \\
\end{eqnarray}

\end{document}